# The effect of surface topography on benthic boundary layer flow: implications for marine larval transport and settlement

Daniel Gysbers[1], Mark A. Levenstein[2], and Gabriel Juarez[3]

[1]*Department of Physics, University of Illinois Urbana-Champaign, Urbana, IL 61801, USA,*

[2]*Université Paris-Saclay, CEA, CNRS, NIMBE, LIONS, 91191, Gif-sur-Yvette, FR, and*

[3]*Department of Mechanical Science and Engineering, University of Illinois Urbana-Champaign, Urbana, IL 61801, USA*

Emails: mark.levenstein@cea.fr, and gjuarez@illinois.edu

**Abstract**

The effect of substrate topography on the settlement of coral larvae in wave-driven oscillatory flow is investigated using computational fluid dynamics coupled to a 2D agent-based simulation of individual larvae. Substrate topography modifies the boundary layer flow by generating vortices within roughness features that can be ejected into the bulk flow, directly influencing larval transport and settlement. In agreement with recent experimental findings, millimeter-scale ridged topographies were found to increase settlement compared to sub-mm feature heights. At this length scale, ridge spacing-to-height ratios of 10 to 20, spacings of more than 30 coral larval body lengths, resulted in the highest settlement rates. These optimal topographies produce a high averaged vertical velocity variance in the bulk flow, indicating that vertical larval movement to benthic surfaces is dominated by passive transport driven by recirculatory flow structures. Indeed, larval settlement was found to be positively correlated with mean vertical velocity variance, and settlement results with substrates comprising complex multiscale roughness were quantitatively similar to those with a simple rectangular model. Our findings reveal how substrates can be designed with surface features to promote larval settlement in natural flow conditions above shallow coral reefs independent of biological cues and substrate material composition.

**Introduction**

  Larval dispersal and settlement are important steps in the life cycle of many sessile marine invertebrates, including environmentally and commercially valuable corals, mussels, and oysters. Adults of these organisms produce sub-millimeter diameter larvae that are dispersed through ocean currents, tides, waves, and other coastal flows to eventually settle in a location in the benthos; hopefully one favorable for survival and growth [1], [2]. However, because of the large spatial and temporal scales of these processes and their relative rarity, it is difficult to observe larval dispersal and settlement in nature [3], [4]. This lack of natural knowledge leads to an unfortunate practical reality: any engineering intervention designed to modify the structure of benthic communities, from *in situ* coral reef restoration to passive anti-fouling measures, is limited by our understanding of the settlement mechanisms and environmental factors that affect larval recruitment.

  An important component of larval dispersal and settlement is the fluid flow near the benthic surface. Coastal ecosystems are reliant on fluid motion for their growth and survival. The fluid motion directly above benthic surfaces, or benthic boundary layer (BBL) flow, influences transport across the surface as well as transport between the water column and the benthic zone. BBL flows are important for the exchange of resources near the surface, from dissolved chemicals [5]–[9] to small particles [10], as well as larval transport and settlement [11], [12]. BBL flows are determined by the surface topography-fluid interaction of the viscous flow profile of water and benthic topography [13]. This interaction can create fluctuating flow structures as oscillatory wave motion passes above complex reef topography, generating velocity gradients and vortices [14], [15]. The effects of these complex flow structures on near surface larval transport are critical to understanding the larval settlement processes.

  Without the ability to study dispersal and settlement processes directly, researchers have relied on field surveys of recruitment patterns [16]–[18], the *in situ* deployment of test settlement substrates [19]–[23], and *ex situ* laboratory experiments [12], [24]–[26]. Numerous surveys and experiments have shown that marine larvae exhibit substrate topography-based settlement preferences [17], [27]–[29]. In particular, larvae of corals, cyprids, bryozoans, and other organisms settle preferentially in topographical features similar or slightly larger than their body size [25], [30]–[33]. This behavior is thought to maximize the contact area between the larvae and the substrate and therefore the number of attachment points (Attachment Point Theory) [24], [25], [33]. However, this theory does not account for flow near surfaces, especially in shallow near-shore environments where hydrodynamic forces dominate larval swimming and attachment [14], [34], [35].

Experiments with larvae in flow have shown that these surface topographies and flow interactions influence larval settlement. Using scleractinian coral as an example, their larvae swim slowly compared to typical reef current speeds, with swimming speeds 1-4 orders of magnitude slower than the flow [12]. This difference can cause larvae to behave similar to passive particles in reef flows. Lower flow speeds generated by velocity fluctuations and eddies that are closer to their swimming speeds are needed for larvae to actively maneuver towards the surface [26], [36]. However, large velocity fluctuations in turbulent flows can also generate shear forces on marine larvae that can prevent them from navigating toward or settling onto a surface [36]. There is thus a tradeoff between flow structures and fluctuations that help larvae settle by reducing flow speeds and others that prevent settlement. While flow-topography interactions in settlement have been studied previously, there has not been a strong focus on characterizing or controlling the surface features or generated flow structures and analyzing how these interact with individual larvae.

We recently performed experiments with coral larvae in oscillatory flow above substrates with controlled topography and found that larvae settled more on substrates with millimeter-scale ridges than on flat substrates with micrometer-scale roughness [37]. Specifically, larvae settled within the gaps created by the ridges, where experimental particle tracking velocimetry (PTV) and computational flow models demonstrated the formation of vortices. Agent-based simulations of larval settlement in these flow models showed that that larvae were transported closer to the surface and trapped in the gaps by the vortices until lower flow speeds achieved in the gaps allowed larvae to swim to the surface and settle. However, only one ridged topography was examined in this study, and topographies of different size and geometry should generate different flow structures with potentially varying effects.

This paper seeks to further our understanding of the connection between surface topography, boundary layer flow, and the settlement of marine larvae. Using a large parameter space that is difficult to access experimentally, we reinforce and extend our previous experimental result and show that, for a given oscillatory flow condition and larval species, there should exist an optimal substrate topography for promoting settlement. Our results demonstrate that the optimal spacing-to-height ratio of this topography results in the creation of not just stationary vortices, but rather of vortices that are ejected out of the roughness feature, thereby increasing the vertical transport of larvae in the water column into substrate refugia. This was confirmed by analyzing time-dependent vertical larval concentration profiles above different topographies and quantified by computing the vertical velocity variance $\langle \sigma_v^2 \rangle$ and correlating it with the larval settlement percentage. Finally, analysis of complex multiscale topographies demonstrated the robust and general character of our results, which can be applied to natural and artificial real-world substrates.

## Methods

*Modelling flow above substrates.* A finite-element model was developed in COMSOL Multiphysics to simulate fluid flow above substrates with different rectangular ridged topographies (**Fig. 1a**). Substrate topographies were varied by ridge height, $h$, from 0.5 to 5 mm and by the spacing between ridges, $w$, from 0.125 to 120 mm. Due to limitations in our computational resources, the largest w/h that could be simulated for substrates with h = 4 and 5 mm was 10 and 4, respectively. Results were compared to a flat substrate, and additionally a selection of multiscale substrates, to compare rectangular ridges to more realistic substrates with complex topography (see Supplementary Information for details). Sections of 2D channels were created in COMSOL with a no-slip bottom wall, top symmetry condition, and periodic side boundaries to simulate an infinite length of repeating ridges. The height of the simulated region was varied between 20 mm and 60 mm to avoid edge effects from ejected vortices (20 mm was the default for computational efficiency unless a taller domain was required). The oscillatory flow was driven by a sinusoidal pressure difference on the periodic side boundaries with a period of 5.5 s and an amplitude dependent on the simulated area to yield a max flow speed of 60 mm s$^{-1}$. These flow conditions were chosen to represent realistic reef flows [37]–[40]. The simulations converged with an average mesh density of 159.36 elements mm$^{-2}$ using a time step of 0.01 s, validated by increasing mesh density and decreasing the timestep until flow velocity profiles were identical [37]. The flow fields produced after 50 periods were utilized to avoid startup effects, and these were imported into MATLAB to simulate individual larval trajectories. Flows that reached periodicity had one period exported that could be looped in the larval simulations. For some substrates with larger $w$, periodicity was not reached even after 100 simulated periods due to large vortices affecting the flow fields of subsequent periods. These flows had similar mean flow speeds between periods with less than a 2.5% difference, but 11 consecutive periods were nonetheless imported into MATLAB to capture the dynamic flow structures operating across several periods.

***Simulating larval transport and settlement.*** Larval trajectories were simulated with a custom 2D agent-based model developed in MATLAB. Larvae were modeled as neutrally buoyant ellipses with a constant swimming velocity, $u_l$, of 3 mm s$^{-1}$ along their major axis (**Fig. 1b**). A swimming speed of 3 mm s$^{-1}$ was chosen as an average of larval swimming speeds from 9 species of corals to represent an idealized coral larva swimmer (**Table S1**) [12]. The larvae were simulated as straight swimmers with no sensing or behavioral response to their environment. Larvae were initiated in a grid of 30 starting positions (three heights of 8, 10, and 12 mm from the surface at 10 equally spaced x-positions above the substrate; $x_i$, $y_i$), 32 initial orientations ($\theta_i$), and 20 initial times equally spaced through the 5.5 s flow period ($\tau$, **Fig. 1a**). Translational motion of the larvae was calculated using the equation:

$$\dot{\boldsymbol{r}} = \boldsymbol{u} + u_l \hat{\boldsymbol{n}},$$

where $\dot{\boldsymbol{r}}$ is the total larval velocity, $\boldsymbol{u}$ is the local flow velocity, and $\hat{\boldsymbol{n}}$ is the unit vector along the larval major axis. Larval rotation was calculated using the equation:

$$\dot{\theta} = \frac{\omega_z}{2} + \alpha \hat{\boldsymbol{g}} \cdot \boldsymbol{E} \hat{\boldsymbol{n}},$$

where $\dot{\theta}$ is the total angular velocity, $\omega_z$ is the vorticity, $\alpha$ is a shape parameter, $\boldsymbol{E}$ is the symmetric rate of strain tensor, and $\hat{\boldsymbol{g}}$ is the unit vector along the larval minor axis. The shape parameter is:

$$\alpha = \frac{1 - \left(\frac{b}{a}\right)^2}{1 + \left(\frac{b}{a}\right)^2},$$

where a is the semimajor axis and b is the semiminor axis (**Fig. 1b**). For a circular larva, $\alpha$ is 0 and the angular velocity is only dependent on vorticity. For a long and narrow larva, $\alpha$ approaches 1, and the angular velocity is primarily dependent on strain as well as vorticity. Here, larvae were simulated with a = 0.25 mm and b = 0.15 mm, which resulted in an $\alpha = 0.47$. In this case, rotation is affected by both vorticity and the rate of strain.

Larvae were seeded into the simulation with an initial position, orientation, and phase in the flow period, and subsequently the translation and rotation equations were used to predict their motion until they settled on a surface, exited from the top of the simulated domain, or more than 10 periods had passed without settlement. Larvae were considered settled when they encountered a surface while their total speed ($\dot{r}$) was lower than the larval swimming speed plus one standard deviation of the average larval speed [37]. Each larva was simulated independently and interactions between larvae were not considered. In the main parametric study, 46 unique conditions were modeled with 19,200 individual larvae trajectories simulated per condition for a total of 883,200 simulated larvae. We additionally simulated 57,600 larval trajectories across three different multiscale ridged topographies.

**Results**

*Larval settlement.* Differences in substrate topography (i.e., $h$ and $w/h$) in the size range investigated have a clear effect on larval settlement percentage (**Fig. 2**). Here, the settlement percentage is defined as the ratio of the number of larvae that settled to the total number of larvae simulated for each condition. For the different topographies studied, the larval settlement percentages range from a minimum of 7% to a maximum of 38%. For all substrates with $w/h \leq 1$, larval settlement is low and independent of ridge height, slightly varying from 7% up to 10%. Furthermore, for substrates with short ridges of $h \leq 1$ mm, settlement remains low, ranging from 7% to 9%, even as $w/h$ increases by an order of magnitude. Conversely, for substrates with tall ridges of $h \geq 2$ mm and $w/h \geq 2$, larval settlement increases significantly and is sensitive to the combination of $h$ and $w/h$. In this region of the phase space, settlement percentage is always greater than 10% and increases up to 38% for a particular combination of $h = 3$ and $w/h = 10$.

Based on these results, emphasis is placed on the substrates with taller ridges that maximize settlement. Specifically, for ridge heights of $h = 2$ mm and 3 mm, larval settlement is non-monotonic with increasing $w/h$ values. In fact, both substrates were found to possess optimum settlement at a particular $w/h$ (**Fig. 2**, star symbol). For the ridge height of $h = 2$ mm, the maximum settlement of 20% was observed for the optimal $w/h = 20$. For $h = 3$ mm, the maximum settlement of 38% was observed for the optimal $w/h = 10$. For comparison, larval settlement onto a flat substrate was simulated and observed to be low, at only 8%. This suggests that the addition of a carefully chosen combination of values of ridge height and $w/h$ can produce up to a five-fold increase in larval settlement for the same flow conditions.

*Flow Structures.* In order to elucidate the above results, we analyzed the transient boundary layer flow structures that developed for each substrate topography. These topographies give rise to three distinct flow regimes during the maximum flow phase of the oscillatory period: skimming flow, un-reattached flow, and reattached flow (**Fig. 3a**, *top row*). The corresponding flow regime for each substrate topography is also indicated on the settlement percent phase map (Fig. 2). In general, settlement percent is low on substrates where skimming flow is observed and high on substrates where un-reattached and reattached flow are observed.

In the skimming flow regime, during both the max and turning point flow phases, vortices are trapped between the ridge gaps and there is little interaction between the bulk flow and the flow within the gaps (**Fig. 3a**, $w/h = 1$). In the un-reattached flow regime, during max flow, vortices do not fill the ridge gaps completely, so streamlines dip down into the gaps. While during the turning point, a single vortex is ejected from between the ridges, enters the bulk flow and quickly dissipates within one half period (**Fig. 3a**, $w/h = 4$). Finally, in the reattached flow regime, during max flow, vortices occur in the ridge gaps, but do not fill the gap region so flow separation, reattachment, and detachment all occur within the gaps. While during the turning point, multiple vortices are ejected high into the bulk flow, extending far above the surface affecting the flow for up to and over a full period (**Fig. 3a**, $w/h = 10$).

*Transport.* Crucially, the behavior of these vortices has a strong influence on larval transport, and thus on settlement. In the skimming flow regime, vertical transport of larvae in the bulk flow is driven by larval swimming alone since the flow is primarily horizontal and no vortices are ejected with which larvae can interact. For un-reattached flow substrates, vortices ejected close above the ridges can transport larvae around them towards the surface. Substrates in the reattached flow regime eject multiple vortices high into the bulk to facilitate larval transport over longer distances.

This is illustrated by examining the movement of larvae above 3 mm-high substrates with a *w/h* ratio of 1, 4, and 10, spanning the three flow regimes, respectively (**Fig. 3b**). After just one flow period, major differences in the transport of larvae from their initial positions are seen. The $w/h = 1$ positions show larvae not mixing in the flow and traveling only by swimming upwards or downwards. Larvae are concentrated above the ridges and few enter the ridge gap. The $w/h = 4$ positions show increased transport near the surface, with larvae entering the ridge gap and settling. Larvae continue to move, solely by swimming, away from the surface at heights larger than 11 mm: the maximum height reached by the ejected vortex. Conversely, a higher settlement substrate of *w/h* = 10 shows larval transport upwards and downwards facilitated by vortices that fill the bulk region and extend to a height of 40 mm. Larvae are concentrated around vortices and just above the surface in the ridge gap.

*Settlement Positions.* The different flow structures present near the substrate surfaces also affect larval settlement positions (**Fig. 3c**). For example, in the skimming flow regime (*w/h* = 1), >35% of larvae settle on the ridge tops rather than inside the gap (1566 total settlers). In the un-reattached flow regime (*w/h* = 4), however, >51% of larvae settle inside the center of the gap (-4 mm < *x* < 4 mm) rather than on ridge tops (4160 total settlers). In the reattached flow regime (*w/h* = 10), >33% of larvae settle inside the gap and concentrated near the ridges (10 mm < |*x*| < 15 mm) (7157 total settlers).

*Transport in bulk region.* To investigate mixing and transport more generally, the vertical velocity variance parameter, $\sigma_v^2$, was computed for all substrate topographies. Vertical velocity variance is a measure of the mean-square deviation from the mean of the vertical velocity, and it has previously been used to investigate flow above reef surfaces and as a measure of the intensity of turbulence [41]–[43]. The parameter is calculated from the equation:

$$\sigma_v^2 = \langle v'v' \rangle,$$

where $v'$ is the fluctuation from the mean vertical velocity.

Time-averaged velocity variance profiles reveal average fluid transport and mixing at each height above the surface during a full period of oscillatory flow (**Fig. 4a**). In the skimming flow regime for *w/h* = 1, $\sigma_v^2$ is zero except at the tops of the ridges and just below them within the gaps, reaching a maximum of 5 mm² s⁻². The low $\sigma_v^2$ in the bulk flow corresponds to low vertical transport by the fluid flow. In the un-reattached flow regime for *w/h* = 4, $\sigma_v^2$ increases to a maximum of 65 mm² s⁻² and the non-zero $\sigma_v^2$ region extends into the bulk flow up to a height of 11 mm. This indicates that there is vertical transport in the flow within four ridge heights of the surface. In the reattached flow regime for $w/h = 10$, $\sigma_v^2$ increases further with a maximum of 150 mm² s⁻² and the non-zero $\sigma_v^2$ region extends far from the surface, up to 58 mm.

Above the optimal $w/h$, $\sigma_v^2$ begins decreasing. For $w/h = 20$, $\sigma_v^2$ decreases to a maximum value of 123 mm² s⁻², and while the non-zero region still extends over 50 mm from the surface, it decreases to less than 1 mm² s⁻² at a height of 40 mm. Finally, for $w/h = 40$, $\sigma_v^2$ decreases further to a maximum value of 55 mm² s⁻² and the non-zero region still extends over 50 mm from the surface.

To directly compare all substrates to each other, accounting for all flow structures and vertical transport, we calculated the $\sigma_v^2$ value within 2 cm of the surface (the minimum height of our model domain). For all substrates, the averaged vertical velocity variance, $\langle\sigma_v^2\rangle$, increases as $w/h$ increases until it reaches a maximum value and then begins to decrease (**Fig. 4b**). When settlement percentage is considered alongside $\langle\sigma_v^2\rangle$, a surprising correlation is revealed. Specifically for substrates with $h = 2$ and 3 mm, the higher settlement substrates corresponded to the larger averaged velocity variance values. The nonparametric Spearman's correlation between $\langle\sigma_v^2\rangle$ and larval settlement percent for all substrate topographies gives a coefficient of 0.663, indicating a positive correlation with a p-value of $p < 0.001$.

*Multiscale topography substrates.* Lastly, to extend our study to more realistic reef surface roughness, flow and larval settlement was simulated for substrates with multiscale topography. Briefly, these substrates were generated by imparting arbitrary small-scale roughness to the rectangular ridged substrates by the summation of sine functions with randomly varying amplitudes (see Supplementary Information for details). The substrate edges were constrained to ensure the periodic flow boundaries were free of any discontinuities and the nominal $h$ and $w/h$ values corresponded to the rectangular ridged substrates. Examples of the multiscale topography substrates are shown for $h = 3$ mm and $w/h = 1$, 4, and 10 (**Fig. 4c**).

Settlement percentages on multiscale topography substrates are very similar to that on their rectangular ridged counterparts (**Table 1**). Specifically, for $w/h = 1$ and 4, the settlement percentages were within <1%, and for $w/h = 10$ the settlement percentages were within <5%. Metrics such as rugosity ($R$) and fractal dimension ($D$) were calculated to compare the roughness between rectangular and multiscale substrates with the same nominal *w/h*, as these are often used for the quantification and assessment of roughness in ecology [44]–[48] (**Table 1** and see Supplementary Information). Despite some differences in roughness metrics, the flow behavior above these multiscale substrates closely matches the behavior seen above rectangular ridged substrates with the same *w/h* ratios. The same flow regimes are found as in the rectangular ridge cases and overall vortices generated by multiscale substrates match both the size and position of vortices above the equivalent rectangular substrates. The vertical velocity variance profiles also reveal the similarities between flow and transport above multiscale and rectangular ridges and allow direct comparison. For each *w/h* value, the $\sigma_v^2$ profiles of multiscale topography substrates have the same shape as their rectangular ridged counterparts and $\sigma_v^2$ values vary by less than 16 mm$^2$ s$^{-2}$ at any height above the surface (**Fig. 4a, dashed lines**). The temporally and spatially averaged $\langle\sigma_v^2\rangle$ values are also similar to those of rectangular ridges, indicating that overall larval transport due to the flow remains the same (**Table 1**).

## **Discussion**

In this paper, we studied the formation of vortices due to surface topography and their effect on marine invertebrate larval settlement. Fluid flow above ridges with different *w/h* ratios produces vortices corresponding to three different flow regimes (skimming flow, un-reattached flow, and reattached flow) that modify larval transport. By examining a large parameter space encompassing the three flow regimes, we investigated how wave-like oscillatory flow influences transport and subsequent settlement trends and identified surface topographies for optimizing settlement hydrodynamics under particular driving flow conditions.

Substrates in the skimming flow region have settlement percentages of less than 10%, comparable to a flat substrate. These substrates have a low value of vertical velocity variance ($\langle\sigma_v^2\rangle$); vortices remain trapped in the ridge gaps, and transport of larvae in the bulk flow is largely unaffected by the substrate. For un-reattached flow substrates, settlement is increased compared to flat substrates and settlement continues to increase with *w/h* up into the reattached flow regime. Vortices ejected from within the ridge gaps travel upwards into the bulk flow until they dissipate, increasing vertical transport above the ridges and leading to an increased $\langle\sigma_v^2\rangle$ value. Finally, reattached flow substrates have high settlement values near the transition from the un-reattached flow regime, but settlement begins to decrease with increasing *w/h*. Multiple vortices that travel over twice as high as in the un-reattached flow regime are ejected at flow turning points, driving larval transport across large areas of the simulations and generating larger $\langle \sigma_v^2 \rangle$ values. Large $\langle \sigma_v^2 \rangle$ values are correlated with higher settlement percentages and can help identify optimal topographies for increasing larval settlement.

The flow regimes remain the same for substrates with multiscale roughness as for substrates with rectangular ridges of the same dominant *w/h* ratio. The smaller scale surface features do not disrupt the large-scale vortices that form and transport larvae to the surface, nor do they create their own flow structures that are relevant at this scale. This suggests that for future studies of the hydrodynamic transport of larvae between the bulk flow and the surface, only millimeter scale features are important to consider and smaller scale roughness can be neglected. Importantly, this demonstrates that these results are relevant to irregular surfaces, which have roughness and imperfections, whether natural or engineered. While near surface interactions will be different and could lead to different settling positions for larvae, the overall transport trends are consistent.

The effects of topography on other aspects of benthic life have been investigated, and topographies of different scales have been found to better facilitate different processes. So-called "bar roughness," like rectangular ridges, can be divided broadly into two types, d- and k-type, based on their *w/h* ratio. When *w/h* is low, usually less than 1 although some studies suggest a more gradual transition for $1 \leq w/h \leq 3$, the roughness is d-type and in the skimming flow regime. When *w/h* is greater than 1, the roughness is k-type and flow is near or beyond the transition to the reattached regime [49]–[53]. For heat and mass transfer to and from the surface, k-type topographies perform better [39]. Larger vortices and greater interaction between the bulk flow and ridge gaps increase not only larval transport, but also the transport of nutrients, oxygen, and waste. However, d-type topographies with smaller ridge gaps shelter larvae better, reducing the drag and pressure experienced by larvae in the gaps [54]. Larvae have also been shown to settle more on surfaces with holes or gaps that are similar to the size of the larvae in still water [30]. According to Attachment Point Theory, larvae can attach more securely to surface features similar to their size by maximizing their number of contact points, or more accurately, the surface area of contact [24], [33].

We found that the optimal topographies for larval settlement are k-type, in the un-reattached and reattached flow regimes. The millimeter scale ridges and gaps investigated here are generally much larger than features predicted to promote settlement according to Attachment Point Theory, but still generate up to a five-fold increase in larval settlement. Conversely, the substrates closest in size to larvae ($h = 0.5$ mm ridged substrates with low $w/h$) have settlement comparable to a flat surface under the oscillatory flow condition studied here. This suggests that while k-type topographies are better suited for the settlement of slow swimming larvae and continued transport between the benthos and the water column, there may be a tradeoff with the benefits provided by other types of substrate refugia. In flow, a combined approach may work better, with larger scale topography to increase transport of larvae to the surface and into gaps and smaller scale features in the more sheltered gaps to allow for better attachment. A detailed study of the long-term effects of different substrate geometries on not only settlement success, but also on recruit survival, is outside the scope of a computational work, but it should be investigated in the future.

In addition to identifying optimal topographies, this study sought to improve metrics for evaluating and designing surfaces for larval settlement. We evaluated the common roughness metrics of rugosity and fractal dimension by comparing rectangular and multiscale substrates and finding which metric best predicted their settlement performance (**Table 1**). Rugosity values were similar for rectangular and multiscale substrates at small $w/h$ and diverged at larger $w/h$, with multiscale substrates having higher rugosity values. The fractal dimension values were different for the two substrate types and relatively unaffected by the nominal $w/h$ ratios. However, neither rugosity nor fractal dimension were correlated with the simulated larval settlement percentage. Conversely, $\langle \sigma_v^2 \rangle$ correlated well with settlement across the simulated $w/h$ range regardless of whether the substrate comprised simple rectangular or complex multiscale topography. This suggests that current roughness metrics alone are not good predictors of settlement and that the flow and its interaction with the topography must be considered. We suggest that $\langle \sigma_v^2 \rangle$ could become a new reference metric, easily calculated from field velocity measurements, to compare and predict the settlement potential of different natural and artificial substrates.

In summary, using a simple numerical model, we have reproduced recent experimental results and demonstrated the existence of surface topographies that produce optimal hydrodynamic conditions for marine larval settlement. These new results have revealed a key mechanism behind the apparent preferential selection of cracks, crevices, and other substrate refugia as settlement locations by slow-swimming larvae who should be otherwise limited in their navigational ability under fast flow conditions. Interactions between oscillatory boundary layer flows and optimal surfaces eject vortices from within suitably sized topographical features, transporting larvae into these features much faster than they could swim on their own. We also identified a new quantitative metric $\langle \sigma_v^2 \rangle$ for evaluating the settlement favorability of substrates under different flow conditions and in designing substrates with optimized millimeter-scale refugia to encourage larval settlement.

**Funding Statement**

This work was supported by the National Science Foundation (IOS 1848671).

**Author Contributions**

D.G., M.A.L., and G.J. conceived of and designed the study; D.G. created and performed the simulations. D.G. analyzed the data and D.G. and M.A.L. wrote the initial manuscript. All authors critically revised the manuscript, gave approval for final publication, and agree to be accountable for all aspects of the work.

**Data Availability Statement**

The datasets generated and/or analyzed during the current study are available from the corresponding author on reasonable request.

**Additional Information**

*Competing Interests.* The authors declare no competing interests.

**Figures and Tables**

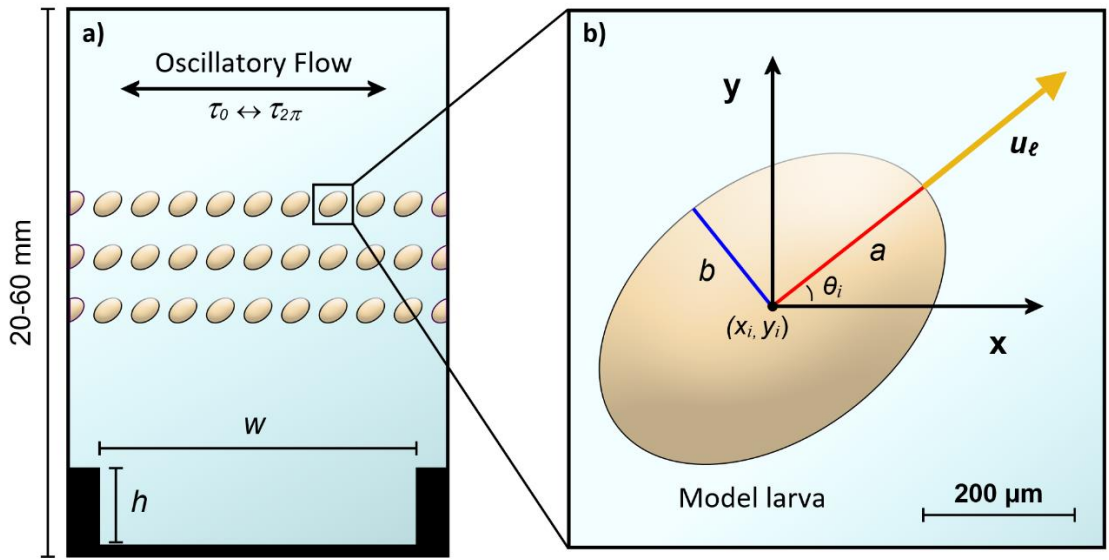

**Figure 1.** Agent-based model of larval settlement. (a) The simulated flow region. Ridge height (*h*) and spacing (*w*) were varied to investigate the effects of substrate topography. Initial larval positions are indicated by beige ellipses, where each larva is simulated independently to avoid larva-larva interactions. The domain height was restricted to 20 mm for computational efficiency, but some simulations were conducted with heights of up to 60 mm to account for vortices ejected high into the bulk flow. (b) A model larva used in settlement simulations with characteristic parameters indicated.

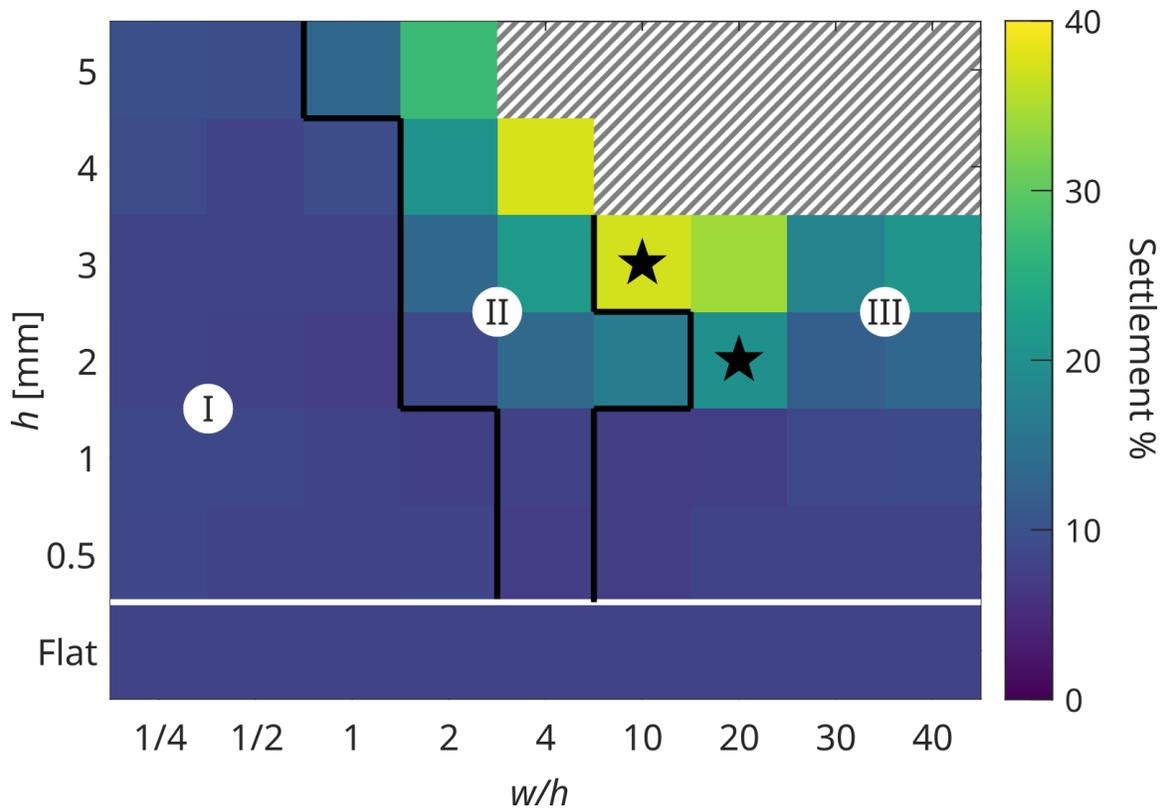

**Figure 2**. Larval settlement percentages for a range of ridge heights and *w/h* values. The stars indicate the highest settlement value, or optimal substrate topography, for each ridge height. Black lines separate substrates by the three observed flow regimes: (I) skimming flow, (II) un-reattached flow, and (III) reattached flow.

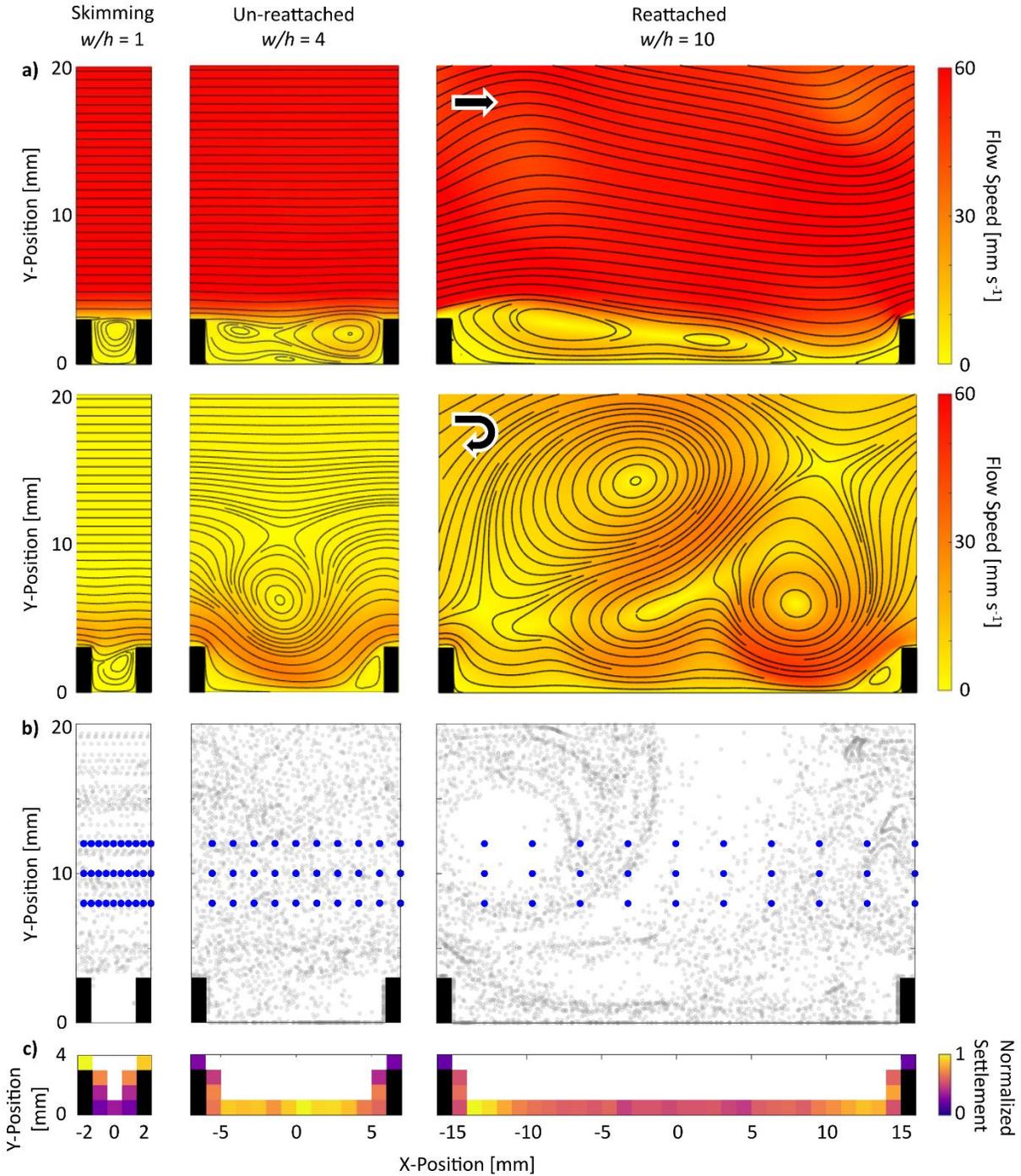

**Figure 3**. Flow structures, larval transport, and settling positions. (a) The simulated flow speed (heatmap) and streamlines (black lines) during the max flow (*top row*) and turning point (*bottom row*) of the oscillatory period for rectangular ridged substrates with $h = 3$ mm *and* $w/h = 1$, 4, and 10. The substrates illustrate the three observed flow regimes: skimming flow regime, un-reattached flow regime, and reattached flow regime. (b) Larval positions above 3 mm tall ridged substrates

for three *w/h* values after one period (5.5 s) of oscillatory flow. Larvae were seeded into the flow at the positions indicated by blue circles. (c) Normalized settlement heatmaps for 3 mm tall ridged substrates for three *w/h* values. Normalization to the maximum settlement bin is performed on each condition independently, where *w/h* = 1 is 8.2% (1566 settlers), *w/h* = 4 is 21.7% (4160 settlers), and *w/h* = 10 is 37.3% (7157 settlers).

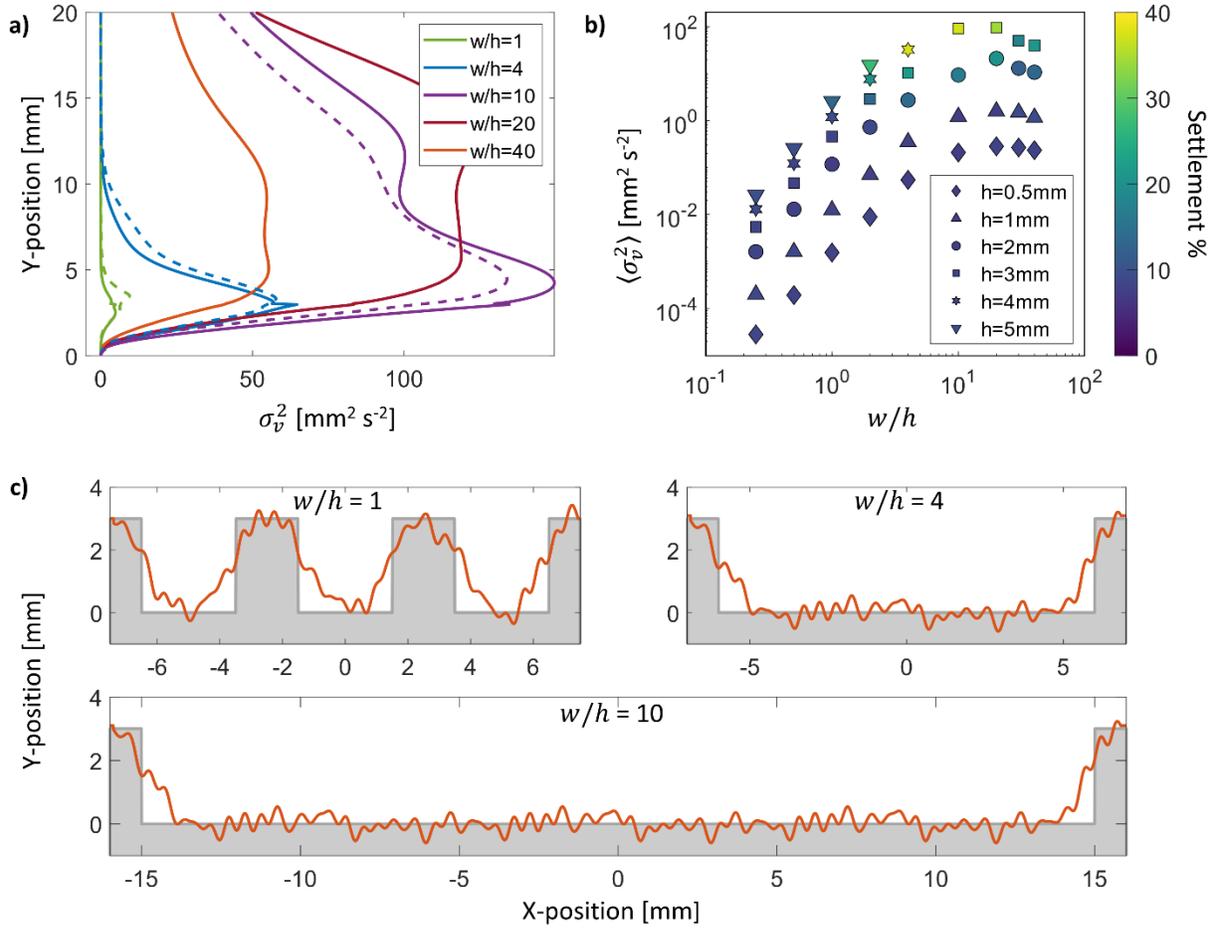

**Figure 4**. Vertical velocity variance analysis and multiscale topography substrates. (a) Temporally and horizontally averaged $\sigma_v^2$ profiles for rectangular ridged substrates with $h = 3$ mm. The profiles for multiscale substrates are indicated by dashed lines. (b) Temporally and spatially averaged $\sigma_v^2$ values plotted against $w/h$ for substrates with 0.5 to 5 mm tall ridges. Data points are colored by larval settlement percentage. For all ridge heights, low $\langle\sigma_v^2\rangle$ substrates display lower settlement percentages and high $\langle\sigma_v^2\rangle$ substrates display high settlement percentages. (c) Comparison of the rectangular ridged substrate and the multiscale topography substrate with a nominal ridge height of 3 mm and nominal $w/h = 1$ (*top left*), $w/h = 4$ (*top right*), and $w/h = 10$ (*bottom*).

| Type | w/h | R | D | $\langle \sigma_v^2 \rangle$ | Settlement % |
|---|---|---|---|---|---|
| Rectangular | 1 | 2.20 | 1.11 | 0.46 | 8.16 |
| Multiscale | 1 | 2.15 | 1.27 | 0.89 | 8.06 |
| Rectangular | 4 | 1.43 | 1.10 | 10.28 | 21.67 |
| Multiscale | 4 | 1.93 | 1.24 | 11.22 | 21.58 |
| Rectangular | 10 | 1.19 | 1.09 | 90.66 | 37.28 |
| Multiscale | 10 | 1.87 | 1.31 | 77.70 | 41.45 |
| Spearman correlation | | -0.77 | -0.09 | 0.83 | - |
| P-value | | 0.10 | 0.92 | 0.06 | - |

**Table 1.** Table comparing rectangular and multiscale topography substrates with $h = 3$ mm and $w/h = 1$, 4, and 10. Metrics shown include the substrate roughness measurements of rugosity ($R$) and fractal dimension ($D$), the temporally and spatially averaged vertical velocity variance $\sigma^2$, and the larval settlement percentage. The nonparametric Spearman's correlation is calculated between the metrics and the settlement percentage.